\documentclass[11pt,twoside]{article}
\usepackage{graphicx}
\usepackage{bm}
\usepackage{subfigure}
\usepackage{multirow}
\usepackage{epstopdf}

\begin{document}

\title{Thin films flowing down inverted substrates: two dimensional flow}
\author{Te-sheng Lin and Lou Kondic\\
\textit{Department of Mathematical Sciences}\\
\textit{and}\\
\textit{Center for Applied Mathematics and Statistics}\\
\textit{New Jersey Institute of Technology, Newark, NJ 07102}}

\date{}

\maketitle

\begin{abstract}
We consider free surface instabilities of films flowing on inverted substrates
within the framework of lubrication approximation.  We allow for the presence of
fronts and related contact lines, and explore the role which they play in
instability development. It is found that a contact line, modeled by a commonly
used precursor film model, leads to free surface instabilities without any
additional natural or excited perturbations.
A single parameter $D=(3Ca)^{1/3}\cot\alpha$, where $Ca$ is the capillary
number and $\alpha$ is the inclination angle, is identified as a governing parameter in
the problem.  This parameter may be interpreted to reflect the combined effect of
inclination angle, film thickness, Reynolds number and the fluid flux.  Variation of $D$
leads to change of the wave-like properties of the instabilities, allowing us to observe
traveling wave behavior, mixed waves, and the waves resembling solitary ones.
\end{abstract}
\maketitle

\renewcommand{\thefootnote}{\fnsymbol{footnote}}

\section{Introduction}

There has been significant amount of theoretical, computational, and experimental work
on the dynamics of thin liquid films flowing under gravity or other body or surface forces
in a variety of settings. The continuous and extensive research efforts are understandable
recalling large number of applications which in one way or another involve dynamics of
thin films on substrates. These applications range from nanoscale assembly, to a variety
of coating applications, or flow on fibers, to mention just a few.

The research activities have evolved in few rather disjoined directions. One of them is
flow down an incline of the films characterized by the presence of fronts (contact lines).
These flows are known to be unstable with respect to transverse instability, leading to
formation of finger-like or triangular patterns~\cite{Hupp82,SD85,CHTC90,deBruyn,BB97}.
One may also consider flow of a continuous stream of fluid down an incline. Experimentally,
this configuration was analyzed first by Kapitsa and Kapitsa~\cite{Kapitsa} and more
recently and in much more detail in a number of works, in particular by Gollub and
collaborators~\cite{Liu_Gollub93,Gollub93,Liu_Gollub94}. The reader is also referred to
~\cite{chang,alekseenko} for relatively recent reviews. Linear stability analysis (LSA)
shows that these films are unstable with respect to long-wave instability when the Reynolds
number, $Re$, is larger than the critical value, $Re_c={5\cot\theta/6}$, where
$\theta$ is the inclination angle~\cite{B1957,yih63}. As the waves' amplitude increases,
LSA cannot describe them anymore as nonlinear effects become dominant.
Therefore, nonlinear models have been developed to analyze this problem, including
Kuramoto-Sivashinsky equation~\cite{Chang94,saprykin_pof05},
Benney equation~\cite{Oron97} and Kapitsa-Shkadov system~\cite{Trifonov91,chang02}.
Typically, film flows exhibit convective instability, suggesting that the shape and
amplitude of the waves are strongly affected by external noise at the source. There
has been a significant amount of research exploring the consequences of imposed
perturbations of controlled forcing frequencies at the inlet~\cite{alekseenko85,
Liu_Gollub94,Malamataris02,nosoko_pof04}. It is found that solitary waves appear at
low frequencies, while saturated sine-like waves occur at high ones. Moreover,
it has been demonstrated that, further downstream, the film flow is dominated by
solitary waves whether they result from imposed perturbations or are `natural'
~\cite{Liu_Gollub94}. It is known that the amplitude and velocity of these waves
are linearly proportional to each other~\cite{alekseenko85,chang93} and the slope
of the amplitude-velocity relation in the case of falling and inclined film had been
examined by several authors~\cite{Liu_Gollub94,kheshgi_pof87,Mudunuri06,Tihon06}.
Another interesting feature is the wave interaction. Since the waves characterized
by larger amplitude move faster, they overtake and absorb the smaller ones.
Furthermore, this merge causes the peak height and velocity to grow significantly
~\cite{Liu_Gollub94,Malamataris02}. In general, the works in this direction
concentrate on a continuous stream of fluid and do not consider issues introduced
by the presence of a contact line.

In a different direction, there is some work on flow of fluids on inverted substrates.
These works involve either the mathematical/computational analysis of the situation
leading to finite time singularity, i.e., detachment of the fluid from the surface
under gravity~\cite{SL00}, or experimental works involving so-called tea-pot
effect~\cite{Pritchard86,Kistler94}, that includes the development of streams
and drops that occur as a liquid film (or parts of it) detach from an inverted
surface, or both~\cite{indeikina97}.
These considerations typically do not include contact line treatment; the fluid film
is assumed to completely cover the considered domain.

In this paper we concentrate on the flow down an inverted inclined substrate of films
with fronts. We will see that this problem includes aspects of all of the rather
disjoined problems considered above. For clarity and simplicity, we will concentrate
here only on two-dimensional flow; therefore we are not going to be concerned with the
three dimensional contact line instabilities. In addition, we assume that the flow
is slow so that the inertial effects can be ignored, and furthermore that the
lubrication approximation is valid, requiring that the gradients of the computed
solutions are small. The final simplification is the one of complete wetting, i.e.,
vanishing contact angle. This final simplification can easily be avoided by including
the possibility of partial wetting, but we do not consider this here. Contact line
itself is modeled using precursor film approach which is particularly appropriate for
the complete wetting case on which we concentrate.  As discussed elsewhere
(e.g.,~\cite{DKB01}), the choice of regularizing mechanism at a contact line does
not influence in any significant way the large scale features of the flow; models
based on any of the slip models will produce results very similar to the ones
presented here.

The main goal of this work is to understand the role of contact line on the formation
of surface waves.  This connection will then set a stage for
analysis of more involved problems, involving contact line stability with respect to
transverse perturbations, and the interconnection between these instabilities.
In addition, the presented research will also allow connecting to the detachment
problem of a fluid `hanging' on an inverted substrate.  We also note that although
we concentrate here on gravity driven flow, our findings with appropriate modifications
may be relevant to the flows driven by other forces (such as electrical or thermal)
and across the scales varying from nano- to macro.

\section{Problem formulation}

Consider a gravity driven flow of incompressible Newtonian film down a planar
surface enclosing an angle $\alpha$ with horizontal ($\alpha$ could be larger than
${\pi/2}$). Assume that the film is perfectly wetting the surface (such as commonly
used silicon oil (PDMS) on glass substrate). Further, assume that lubrication
approximation is appropriate, as discussed in~\cite{siam}. Within this approach,
one finds the following result for the depth averaged velocity ${\bar {\bf v}}$
\begin{equation}
3 \mu \bar{\bf v}=\gamma \bar{h}^2 \bar{\nabla} \bar{\nabla}^2 \bar{h} -
\rho g \bar{h}^2\bar{\nabla} \bar{h}\cos\alpha+ \rho g \bar{h}^2\sin\alpha\bf{i}\ ,
\end{equation}
where $\mu$ is the viscosity, $\gamma$ is the surface tension, $\rho$ is the density,
$g$ is the gravity, $\bar{h}=\bar{h}(\bar{x},\bar{y},\bar{t})$ is the fluid thickness
and $\bar{\nabla} = (\partial_{\bar{x}},\partial_{\bar{y}})$ ($\bar{x}$ points
downwards and $\bar{y}$ is in horizontal transverse direction). By using this expression
in the mass conservation equation, ${\partial \bar{h}}/{\partial \bar{t}}+\bar{\nabla}
\cdot \left( \bar{h}\bar{\bf v}\right) =0$,  we obtain the following dimensionless PDE
~\cite{schwartz89,oron92}
\begin{equation}
\frac{\partial h}{\partial t}+\nabla \cdot \left[ h^3\nabla \nabla^2
h\right] - D \nabla\cdot[h^3\nabla h]
+ \frac{\partial h^3}{\partial x}=0\ . \label{eq:dif-g}
\end{equation}
Here, thickness $h$ and coordinates $x$, $y$ are expressed in units of
$h_0$ (the fluid thickness far behind the front), and $\ell =h_0
(3Ca)^{-1/3}$, respectively.  The capillary number,  $Ca={\mu U/\gamma}$,
is defined in terms of the flow velocity $U$ far behind the front.
The time scale is chosen as $\ell /U$.  Single  dimensionless parameter
$D=(3Ca)^{1/3}\cot\alpha$ measures the size of the normal component of gravity.

Equation~(\ref{eq:dif-g}) requires appropriate boundary and initial conditions,
which are formulated below.   We concentrate on the physical problem where uniform
stream of fluid is flowing down an incline and therefore far behind the fluid front
we will assume the fluid thickness to be constant (constant flux configuration).
At the contact line itself, we will assume `precursor film model' discussed in some
detail elsewhere~\cite{DKB01}. The initial condition is put together with the idea of
modeling incoming stream of the fluid, and the only requirement is that it is consistent
with the boundary conditions.

For the flow down an inclined surface with $\alpha \le {\pi/2}$ and correspondingly
$D\ge 0$, the solutions of Eq.~(\ref{eq:dif-g}) are fairly well understood both in the two
dimensional (2D) setting where $h=h(x,t)$, and in the 3D one, where $h=h(x,y,t)$. The
solution is characterized by a capillary ridge which forms just behind the front~\cite{KD02}.
This solution, while stable in 2D, is known to be unstable to perturbations in the
transverse, $y$, direction. In this work, we will concentrate on the 2D setup only,
but for $D<0$, therefore analyzing the flow down an inverted surface (hanging film).

\subsection{Initial and boundary conditions}

Consider two dimensional flow, therefore $h$ is $y$-independent. Equation~(\ref{eq:dif-g})
can be rewritten as
\begin{equation}
\frac{\partial h}{\partial t}+\left[ h^3 \left(h_{xxx}- D h_x+1\right)\right]_{x} =0 \ .
\label{eq:1D}
\end{equation}
The numerical simulation of Eq.~(\ref{eq:1D}) is performed via a finite-difference method.
More specifically, we have implemented implicit 2nd-order Crank-Nicolson method in time,
2nd-order discretization in space and Newton's method to solve the nonlinear system
in each time step, as described in detail in e.g.,~\cite{siam}. The boundary conditions
are such that constant flux at the inlet is maintained. The choice implemented here is
\begin{equation}
h(0,t)=1, \quad h_{xxx}(0,t)-Dh_x(0,t)=0.
\end{equation}
At $x = L$,  we assume that the film thickness is equal to the precursor, so that
\begin{equation}
h(L,t)=b, \quad h_x(L,t)=0,
\end{equation}
where $L$ is the domain size and $b$ is the precursor film thickness, $b\ll 1$.
Typically, we set $b=0.01$. The initial condition is chosen as a hyperbolic tangent
to connect smoothly $h=1$ and $h=b$ at $x=x_f$; it has been verified that the results are
independent of the details of this procedure.

\section{\label{sec:comp}Computational results}

It is known that for flow down a vertical plane, a capillary ridge forms immediately
behind the fluid front. This  capillary ridge can be thought of as a strongly damped
wave in the streamwise direction. As we will see below, this wave is crucial for
understanding the instability that develops for a flow down an inverted surface. Here,
we first outline the results obtained for various $D$'s, and then discuss their main
features in some more details in the following section.  We use $x_f = 5$ for all
the simulations presented in this section.

\begin{figure}[th]
\centering
\includegraphics[scale=0.32]{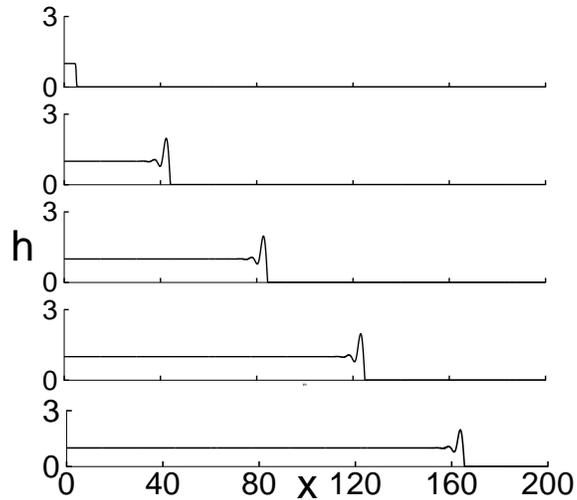}
\caption{The flow down an inverted substrate ($D = -1.0$). From top to bottom, $t=0$, $40$,
$80$, $120$, $160$.}
\label{fig:d10}
\end{figure}

{\it Type 1: $-1.1\leq D< 0$.}  For these values of $D$, we still observe existence of
a dominant capillary ridge; this ridge becomes more pronounced as the magnitude of $D$
is increased.  In addition, we also observe secondary, strongly damped oscillation behind
the main ridge. Figure~\ref{fig:d10} shows an example of time evolution profile for
$D=-1.0$.   For longer times, traveling wave solution is reached, and the wave speed
reaches a constant value equal to $U = 1+b+b^2$, as discussed, e.g., in~\cite{BB97}.
Appendix~\ref{sec:tws} gives more details regarding this traveling wave solution, including
discussion of the influence of precursor film thickness on the results, see
Fig.~\ref{fig:tail_tw}.

\begin{figure}[th]
\centering
\includegraphics[scale=0.32]{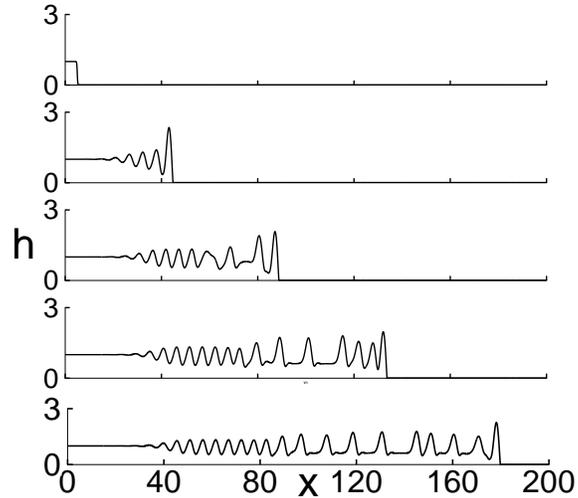}
\caption{The flow down an inverted substrate ($D = -1.5$). From top to bottom, $t=0$,
$40$, $80$, $120$, $160$ (enhanced online).}
\label{fig:d15}
\end{figure}

\begin{figure}[th]
\centering
\includegraphics[scale=0.38]{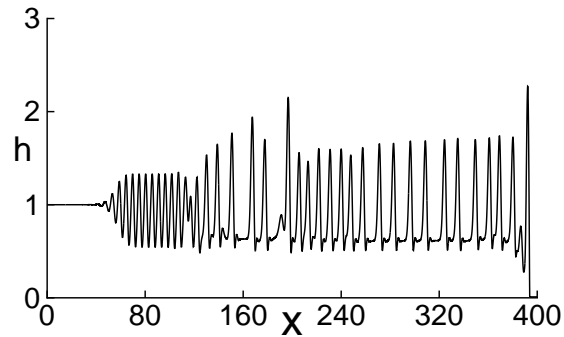}
\caption{The flow down an inverted substrate ($D = -1.5$) at $t=340$.}
\label{fig:d15_340}
\end{figure}

{\it Type 2: $-1.9 \leq D < -1.1$}.  The capillary ridge is still observed; however,
here it is followed by a wave train. Figure~\ref{fig:d15} shows as an example of the
evolution for $D=-1.5$.  Waves keep forming behind the front, and, furthermore, they
move {\it faster} than the front itself. Therefore the first wave behind the front
catches up with the ridge, interacts and merges with it. The other important feature
of the results is that there are three different states observed behind the capillary
ridge: two types of waves and a constant state.  These states can be clearly seen in
the last frame of Fig.~\ref{fig:d15}. Immediately behind the front, there is a range
characterized by waves resembling solitary ones~\cite{chang} discussed in some more
detail below. This range is followed by another one with sinusoidal shape waves. Finally
there is a constant state behind. Such mixed-wave feature remains present even for
very long time. To illustrate this, Fig.~\ref{fig:d15_340} shows the result at much
later time, $t=340$, using an increased domain size.

\begin{figure}[h]
\centering
\includegraphics[scale=0.39]{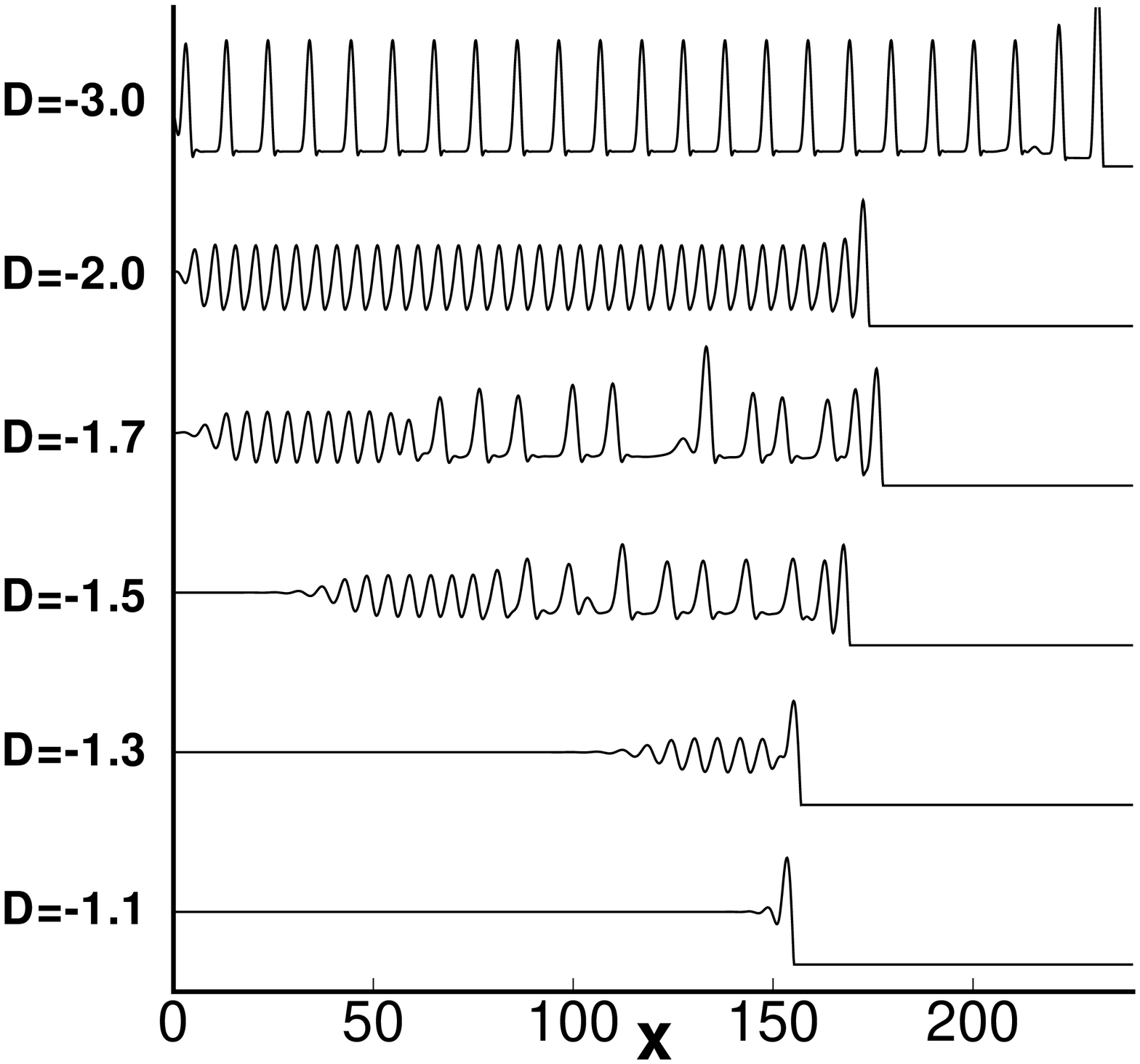}
\caption{Comparison of the results for different $D$'s at $t=150$.}
\label{fig:D_compare}
\end{figure}

Also, Fig.~\ref{fig:D_compare}, which includes typical results from the {\it Type 3}
regime discussed below, imply that {\it Type 2} corresponds to  a transitional regime
between the  {\it Types 1} and {\it 3}.  Additional simulations (not shown here) suggest
that the regions where the waves are present within {\it Type 2} regime become more and
more extended as the magnitude of $D$ is increased. Future insight regarding the nature
of wave formation in {\it Type 2} regime is discussed in the following Section. Here we
note that the available animations of wave evolution are very helpful to illustrate the
complexity of wave interaction in {\it Type 2} and {\it Type 3} regime discussed next.

\begin{figure}[th]
\centering
\includegraphics[scale=0.32]{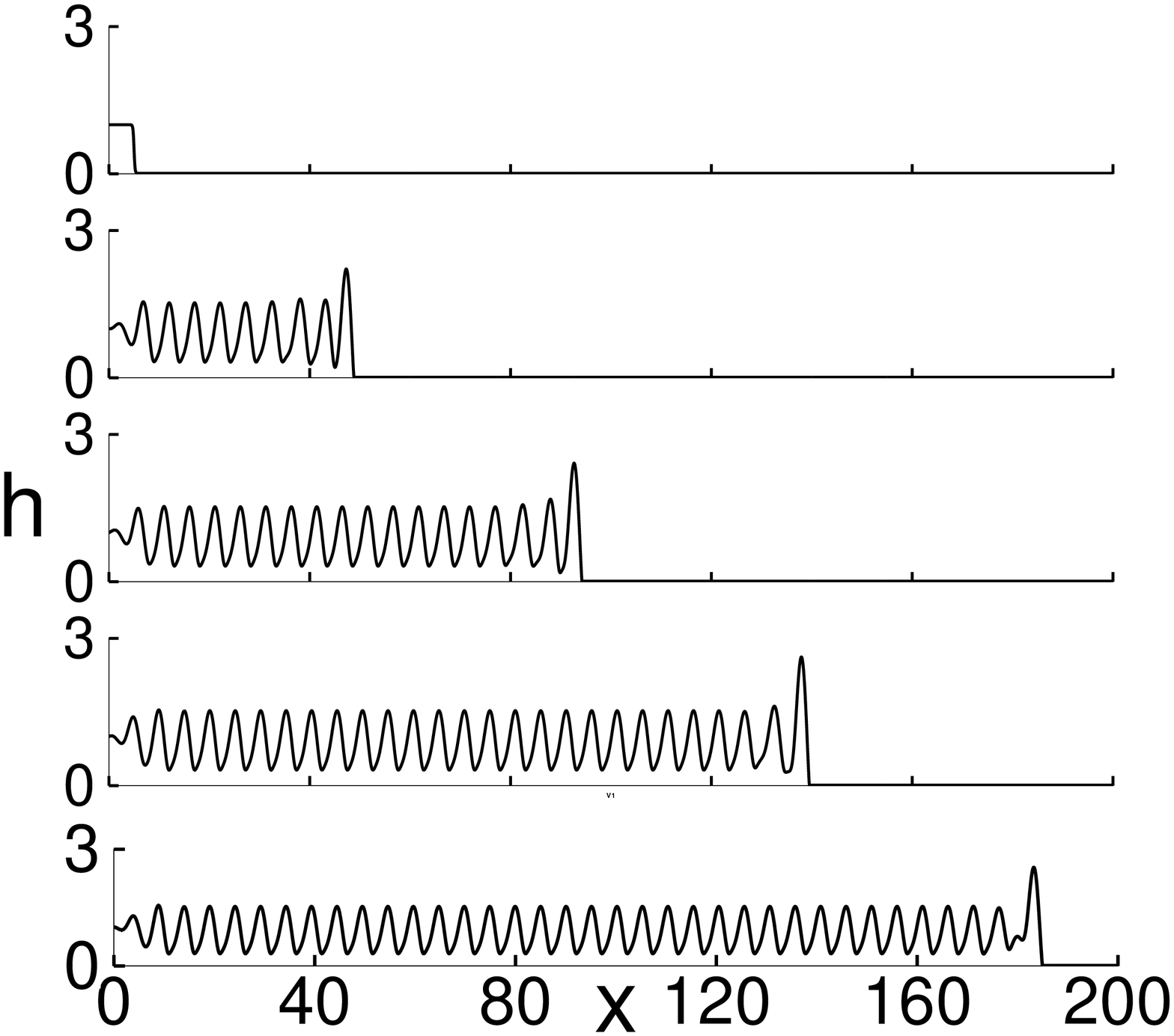}
\caption{The flow down an inverted substrate ($D = -2.0$). From top to bottom,
$t=0$, $40$, $80$, $120$, $160$. Note that there is a continuous interaction of the
surface waves and the front, since the surface waves travel faster than the front
itself (enhanced online).
}
\label{fig:d20}
\end{figure}

{\it Type 3:  $-3.0\leq D<-1.9$}.   This is a nonlinear steady traveling wave regime.
There is no damping of surface oscillations that we observed e.g., in Fig.~\ref{fig:d15}.
Figure~\ref{fig:d20} shows an example obtained using $D=-2.0$. Here, a wave train forms
behind the first (still dominant) capillary ridge. Similarly as before, since this wave
train travels faster than the fluid front, there is an interaction between the first of
these waves and the capillary ridge.  On the other end of the domain, these waves
also interact with the inlet at $x=0$; the role of this interaction is discussed in more
details later in Sec.~\ref{sec:C}.

\begin{figure}[th]
\centering
\includegraphics[scale=0.30]{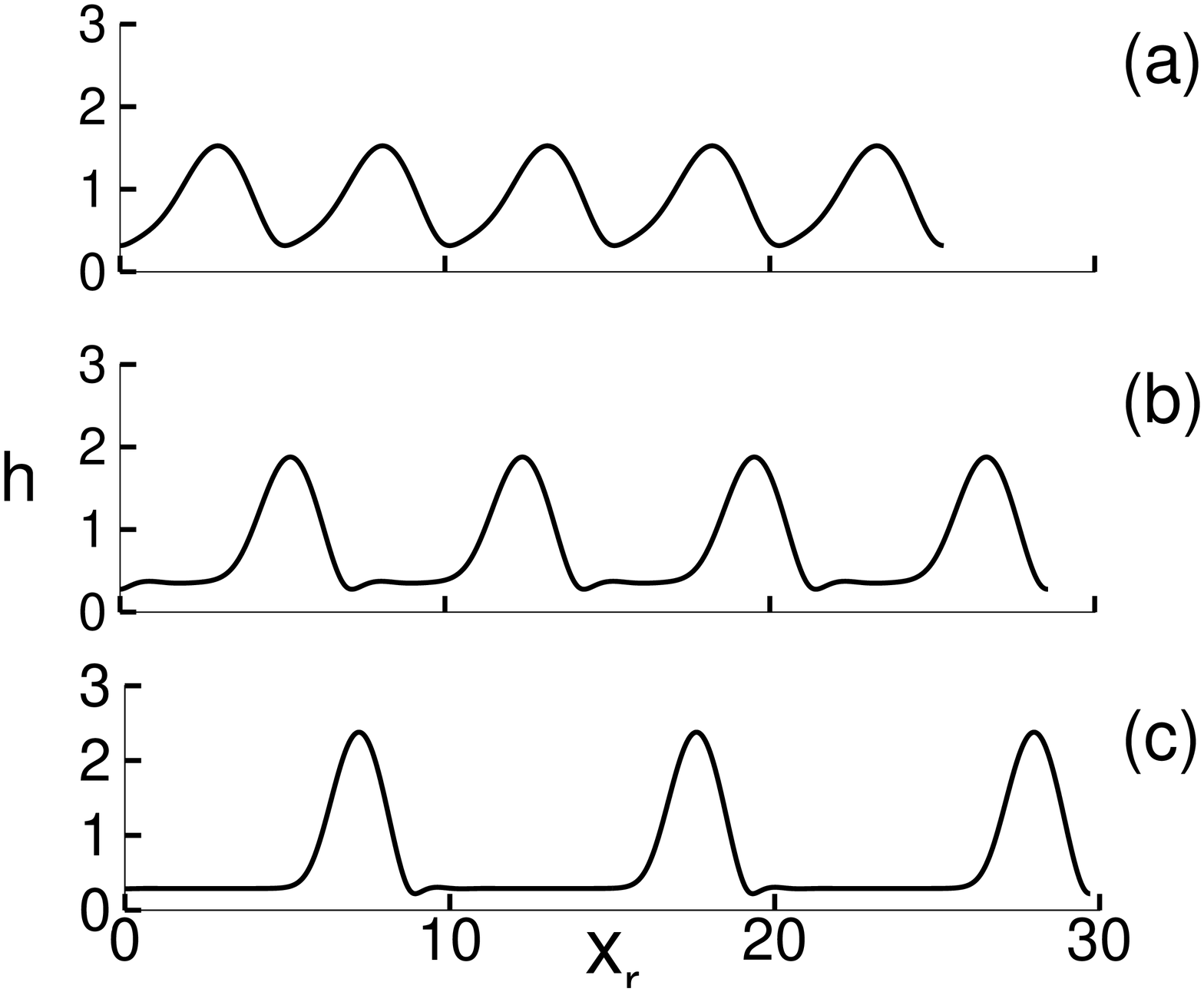}
\caption{Wave profile for different $D$'s. (a): $D=-2.0$; (b): $D=-2.5$; (c):
$D=-3.0$. The wave profile have been shifted to illustrate the difference in
wave number, i.e., $x_r=x-x_0$ where $x_0$ is an arbitrary shift.}
\label{fig:wave}
\end{figure}

We find that the {\it Type 3} includes two sub-types. For smaller absolute values of $D$,
such as $D= -2.0$,  one finds sinusoidal waves as shown in Fig.~\ref{fig:d20}. For larger
magnitudes of $D$, we find solitary type waves, the structures  sometimes referred to as
`solitary humps', such that characteristic dimension of a hump is much smaller than the
distance between them~\cite{chang}. Both types of waves are illustrated in
Fig.~\ref{fig:D_compare}, which shows the results for $D=-2.0$ and $D=-3.0$, and in
Fig.~\ref{fig:wave}, showing the typical wave profiles for $D=-2.0$, $D=-2.5$ and $D=-3.0$.
The wave profiles that we find are very similar to the ones observed for continuous films
exposed to periodic forcing~\cite{Liu_Gollub94,chang93,nosoko_pof04}.   For the flow
considered here, the governing parameter is $D$, in contrast to the forcing frequency in
the works referenced above.

In the next section, we will discuss in more detail some features of the results presented
here. Here, we only note that it may be surprising that all the waves discussed are found
using numerical simulations, remembering that we do not impose forcing on the inlet region,
and furthermore, we do not include inertial effects in our formulation. Instead, we have a
hanging film with a contact line in the front. Therefore, it appears that the presence of
fronts and corresponding contact lines plays an important role in instability development.

We note that it is possible in principle to carry out the computations also for more
negative $D$'s. We find that, as absolute value of $D$ is increased, the amplitude of
the waves, including the capillary ridge, increases, and furthermore, the periodicity
of the wave train following the capillary ridge is lost. However, since the observed
structures are characterized by relatively large spatial gradients which at least locally
are not consistent with the lubrication approximation, we do not show them here. It
would be of interest to consider this flow configuration outside of lubrication
approximation and analyze in more detail the waves in this regime. In addition, this
regime should also include the transition from flow to detachment, the configuration
related to the so call `tea-pot' effect~\cite{Pritchard86,Kistler94,indeikina97}.

\section{Discussion of the results}

In this Section, we discuss in some more detail the main features of the numerical
results and compare them to the ones that can be found in the literature. We consider
in particular the difference between various regimes discussed above. In Sec.~\ref{sec:A}
we give the main results for the velocities of the film front and the propagating waves.
In Sec.~\ref{sec:B} we discuss the main features of the instability that forms and
show that the presence of contact line is important in determining the properties
of the waves, including their typical wavelength. Then we finally discuss one question
that was not considered explicitly so far: What is the source of instability? As we
already suggested, contact line appears to play a role here.  However, it is appropriate
to also discuss the influence of numerical noise on instability development, shown in
Sec.~\ref{sec:C}. As we will see, both aspects are important to gain better understanding
of the problem.

\subsection{\label{sec:A}Front speed and wave speed}

\begin{figure}[th]
\centering
\includegraphics[scale=0.35]{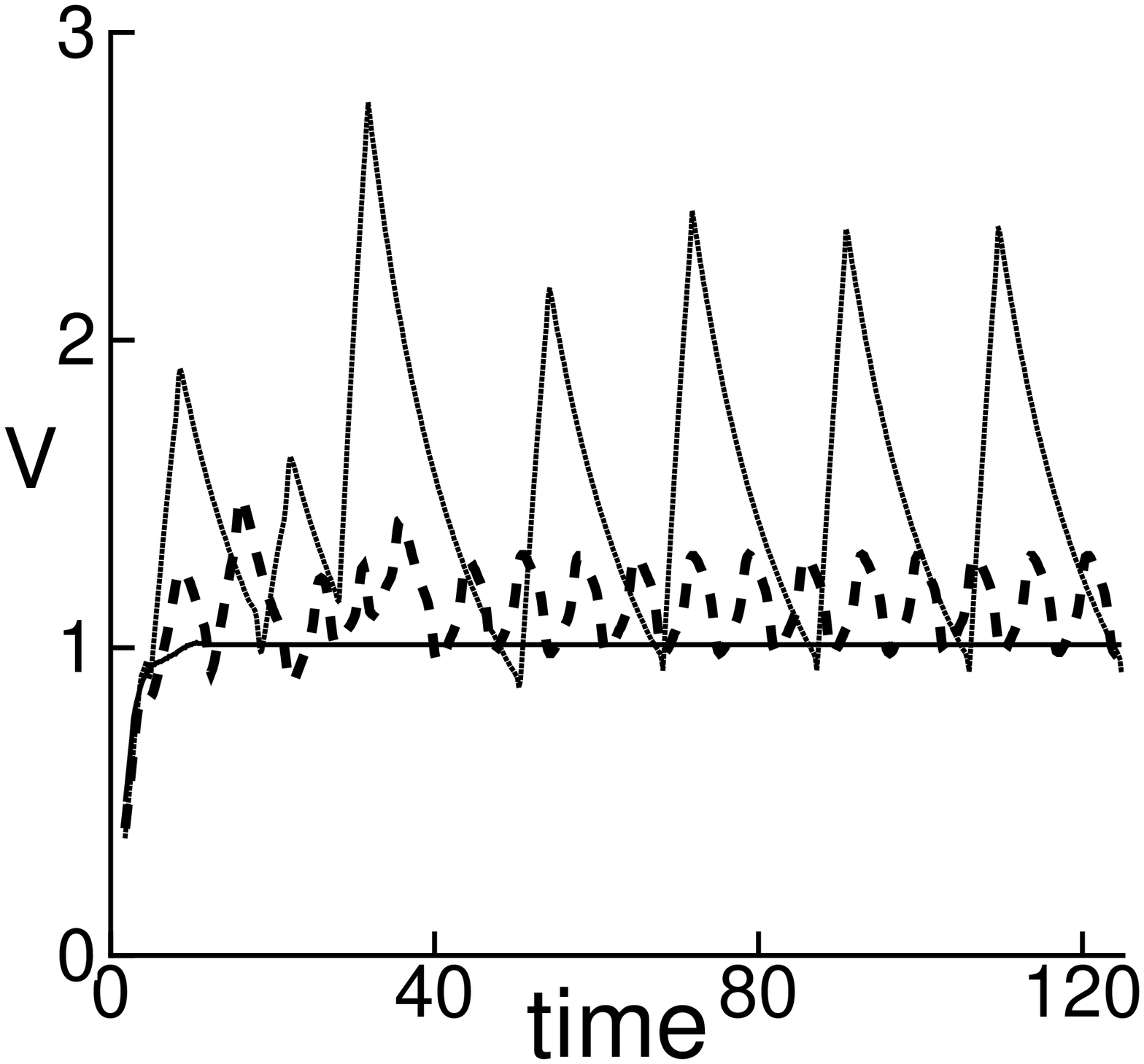}
\caption{Velocity profile of the leading capillary ridge for different $D$'s. $D=-1.0$
(solid), $D=-2.0$ (dashed), $D=-3.0$ (dotted).}
\label{fig:velocity}
\end{figure}

Figure~\ref{fig:velocity} compares the velocity of the leading capillary ridge for
different $D$'s. The speed of the traveling wave solution, $U$, is $1+b+b^2$, and is
exactly the front speed for $D = -1.0$, as discussed in Appendix~\ref{sec:tws} in
connection to Fig.~\ref{fig:tail_tw}. For all other cases shown, the velocity of the
leading capillary ridge oscillates around $U$, due to the interaction between the
leading capillary ridge and the upcoming waves.

\begin{figure}[th]
\centering
\includegraphics[scale=0.40]{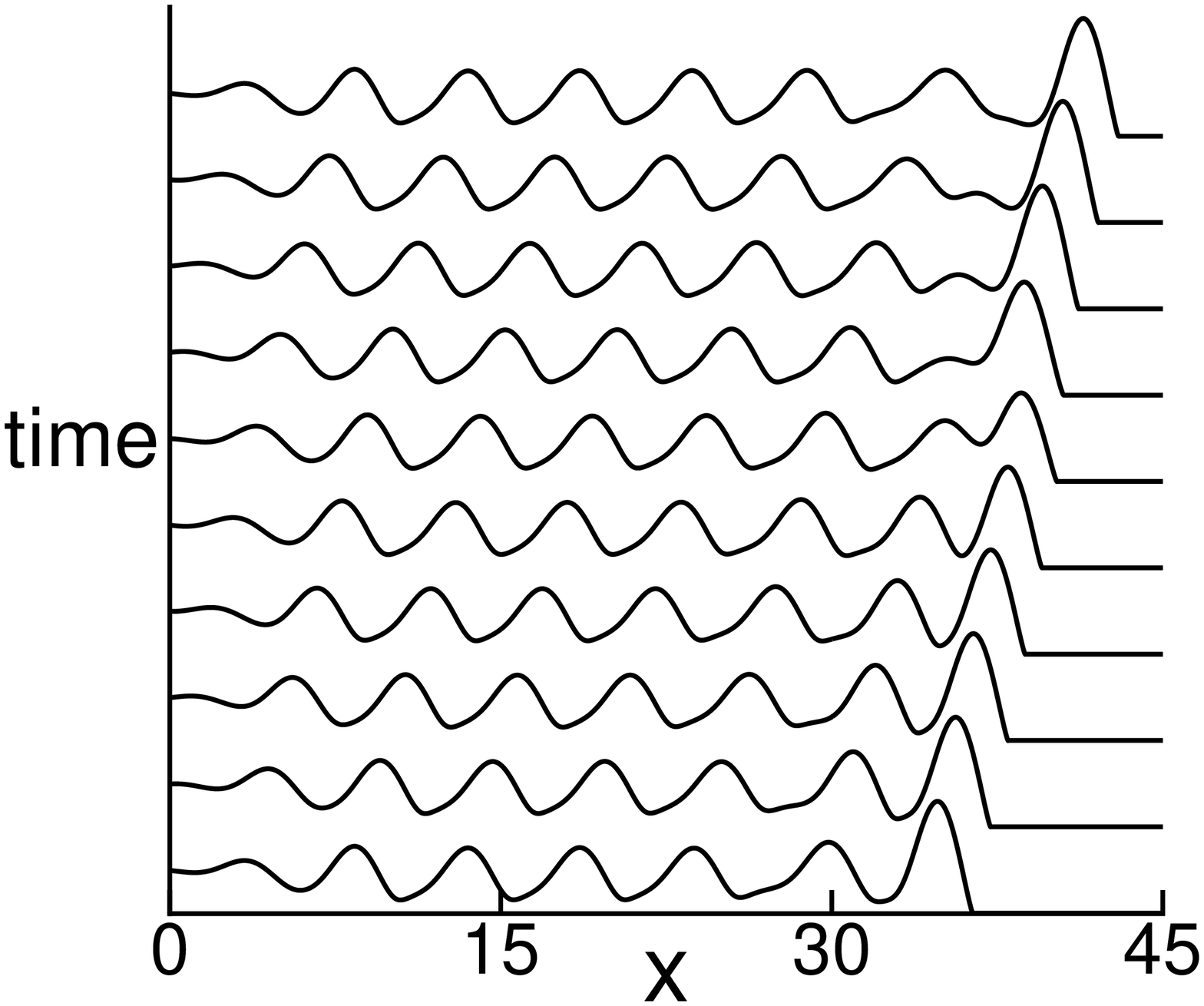}
\caption{Wave interaction with the capillary ridge. Time evolution is from the bottom to the
top, $D=-2.0$.}
\label{fig:evol}
\end{figure}

Table~\ref{tab:speed3} shows the speed of waves in {\it Type 3} regime. As we can see,
the wave speed in all cases is greater than $U$. A simple explanation on why the waves
move faster than the fluid front itself is that the motion of the front is resisted by the
precursor film (recall that in the limit $b\rightarrow 0$, there is infinite resistance to
the fluid motion within the formalism implemented here). The surface waves however, travel
with a different, larger speed. Therefore, the upcoming waves eventually catch up with the
front, interact, and merge into a new capillary ridge. Figure~\ref{fig:evol} illustrates this
process. Due to the conservation of mass, the height of the leading capillary ridge increases
strongly right after the merge. Also the speed of the capillary ridge increases. As the
leading capillary ridge moves forward, its height decreases until next wave arrives. That
is the reason why we see such pulse-like velocity profiles in Fig.~\ref{fig:velocity}; each
pulse is a sign of a wave reaching the front. In the {\it Type 2} regime, the velocity of the
front shows similar oscillatory behavior, although the approximate periodicity of the
oscillations is lost due to more irregular structure of the surface waves. Going back to
the {\it Type 3} and Table~\ref{tab:speed3}, we see that wave amplitude and speed are both
increasing with $D$, consistently with the behavior of continuous vertically falling
films~\cite{alekseenko85,chang93}.

\begin{table}[th]
\centering
\renewcommand{\tabcolsep}{.4cm}
\begin{tabular}[t]{ccc}
\hline
 $D$ &  wave amplitude & wave speed\\
\hline
-2.0 & 1.53 & 1.88\\
-2.5 & 1.88 & 2.20\\
-3.0 & 2.38 & 2.65\\
\hline
\end{tabular}
\caption{Wave amplitude and wave speed for different $D$'s in {\it Type 3} regime.
Note that the wave speed is always larger than the capillary ridge speed, shown in
Fig.~\ref{fig:velocity}.}
\label{tab:speed3}
\end{table}

\subsection{\label{sec:B}Absolute versus convective instability}

Here we analyze some features of the results in particular from {\it Type 1} and
{\it Type 2} regimes using linear stability analysis. Let us ignore for a moment
the contact line, and analyze stability of a flat film. The basic framework is given
in the Appendix A2. We realize that Eq.~(\ref{linear}) can be reduced to a linear
Kuramoto-Sivashinsky equation in the reference frame moving with the nondimensional
speed equal to $3$. Consider then the evolution of a localized disturbance imposed
on the flat film at $t=0$. This disturbance will transform into an expanding wave
packet with two boundaries moving with the velocities $(x/t)_-$ and
$(x/t)_+$~\cite{huerre90}. In the laboratory frame, these velocities are given
by~\cite{chang95}
\begin{equation}
{\left( x \over t\right)}_\pm \approx 3 \pm 1.62(-D)^{3/2}~.
\label{eq:lin}
\end{equation}
The right going boundary moves faster than the capillary ridge and can
be ignored. Considering now the left boundary, we see that there is a range
$D_{c1} > D > D_{c2}$ such that the speed of this boundary is positive and
smaller than $U$. Alternatively, one can use the approach from~\cite{focas05},
which is based on studying the behavior of the curve $\omega_i=0$ in the
complex $k$ plane, with the same result. Using either approach, one finds
$D_{c1}  \approx - 1.15$ and $D_{c2} \approx - 1.51$. This result explains
the boundary between the {\it Type 1} and {\it Type 2} regimes since
for {\it Type 1}, $D> D_{c1}$ and the left boundary moves faster than the
front itself.  For $D > D_{c2}$, the speed of the left boundary is positive,
and therefore the instability is of convective type. This can be seen from
Fig.~\ref{fig:D_compare} and is illustrated in detail in Table~\ref{tab:Dc}.
In this table, we show the value of $D_{lin}$, predicted by Eq.~(\ref{eq:lin}),
using the position of the left boundary $x_-$ obtained numerically.
While the agreement between $D_{num}$ and $D_{lin}$ is generally
very good, we notice some discrepancy for $D_{num} = -1.7$; this can
be explained by the fact that for this $D_{num}$ there is already some
interaction with the boundary at $x=0$.

These results suggests that we should split our {\it Type 2} regime into two parts:
{\it Type 2a}, for which the speed of the left boundary  is positive
($D > D_{c2}$), and {\it Type 2b}, for which the speed of the boundary
is negative and the instability is of absolute type. In {\it Type 2a} regime,
a flat film always exists and expands to the right with time. In {\it Type 2b}
regime, flat film disappears after sufficiently long time. As an illustration,
we note that $D=-1.5$ shown in Fig.~\ref{fig:d15}, lies approximately at the
boundary of these two regimes, since here the length of the flat film is almost
time-independent. We also note that in {\it Type 2b} regime, we always
observe two types of waves, in contrast to {\it Type 3}; that is, the structure
shown e.g. in Fig.~\ref{fig:D_compare} for $D=-1.7$ persists for a long time.

\begin{table}[th]
\centering
\renewcommand{\tabcolsep}{.3cm}
\begin{tabular}[t]{cccc}
\hline
$D_{num}$ &$x_-$ & $(x/t)_-$ & $D_{lin}$ \\
\hline
-1.1 & 150 & 1.00 &  -1.15\\
-1.3 & 100 & 0.66 &  -1.28\\
-1.5 & 30  & 0.20 &  -1.44\\
-1.7 & 0   & 0.00 &  -1.51\\
\hline
\end{tabular}
\caption{$D_{num}$ is the value of $D$ used in simulations, the position $x_-$ is
taken from Fig.~\ref{fig:D_compare} ($t=150$) and used to calculate the speed
$(x/t)_-$; $D_{lin}$ is calculated from Eq.~(\ref{eq:lin}) using the negative sign.}
\label{tab:Dc}
\end{table}

\begin{figure}[th]
\centering
\includegraphics[scale=0.32]{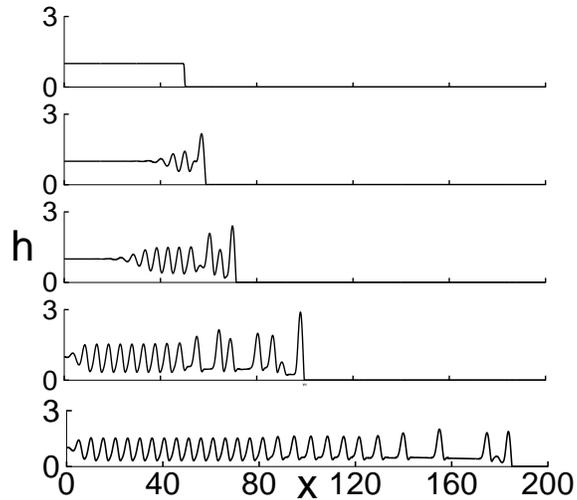}
\caption{$D=-2.0$. From top to bottom, $t=0$, $10$, $20$, $40$, $100$. For
early times, the contact line induced instability propagates to the left. For longer
times, sine-like and solitary-like waves are observed, covering the whole domain
by $t=40$ (enhanced online).
}
\label{fig:d20102}
\end{figure}

To allow for better understanding of the properties of the waves that form, in the
results that follow we have modified our initial condition (put $x_f = 50$) to allow
for longer wave evolution without interaction of the wave structure with the domain
boundary ($x=0$). Figure~\ref{fig:d20102} shows that for $D=-2.0$, the waves form
immediately behind the leading capillary ridge; see also the animation attached to
this figure. For longer time ($t>40$ in Fig.~\ref{fig:d20102}), the disturbed region
covers the whole domain as expected based on the material discussed in Sec.~\ref{sec:B}.
Note that even for $t=100$ we still see transient behavior: the long time solution for
this $D$ consists of uniform stream of waves and is shown in Fig.~\ref{fig:wave}(a).
This long time solution is independent of the initial condition. However, the time
period needed for this uniform stream of waves to be reached depends on the initial
film length and is much longer for larger $x_f$ used here.

Figure~\ref{fig:d20102} suggests that contact line plays a role in wave formation
(the other candidate, numerical noise, is discussed below). One may think of contact
line as a local disturbance. It generates an expanding wave packet as we have just
shown, and the velocities of the two boundaries are given by Eq.~(\ref{eq:lin}). In
particular, for $D < D_{c1}$, the left boundary moves slower than the capillary ridge.
The wave number, $k_l$, along this boundary is defined by
\begin{equation}
\label{eq:kl}
\left. \left(\frac{\partial \omega}{\partial k}\right) \right|_{k=k_l}
=\left(\frac{x}{t}\right)_-,
\end{equation}
and it should be compared  to the sine-like waves that form due to contact line presence
(such as the  waves shown in Fig.~\ref{fig:d20102} for early times). Table~\ref{tab:wavelength}
gives this comparison: the values of $k_l$ for a given $D$ are shown in the second column,
followed by the numerical results for the wave number, $q_n$. We find close agreement,
suggesting that $k_l$ captures very well the basic features of the waves that form due
to contact line presence. Furthermore, both $q_n$ and $k_l$ are {\it much larger} than
$k_m$, the most unstable wave number expected from the linear stability analysis (LSA)
described below (vis. the last column in Table~\ref{tab:wavelength}). This difference
allows to clearly distinguish between the contact line induced waves and the noise induces
ones, discussed in what follows.

\begin{table}[th]
\centering
\renewcommand{\tabcolsep}{.3cm}
\begin{tabular}[t]{cccc}
\hline
$D$ & $k_l$ & $q_n$ &  $k_m$ \\
\hline
-1.3 & 1.10&  1.10 &  0.81\\
-1.5 & 1.18&  1.14 &  0.87\\
-1.7 & 1.25&  1.23 &  0.92\\
-2.0 & 1.36&  1.35 &  1.00\\
-2.5 & 1.52&  1.48 &  1.12\\
\hline
\end{tabular}
\caption{
The second column shows theoretical results for the wave number of the left moving
boundary ($k_l$), in the limit of small oscillations, see Eq.~(\ref{eq:kl}). The third
column shows the wave number resulting from simulations for different $D$'s in {\it Type 2}
and {\it Type 3} regimes in the contact line induced part (e.g., the waves shown in
Fig.~\ref{fig:d15} ($t=160$) for $40< x < 80$ for {\it Type 2}, or in Fig.~\ref{fig:d20102}
($t=20$) about $x=40$). The last column shows
the wave number of maximum growth, $k_m$, resulting from the linear stability analysis.
}
\label{tab:wavelength}
\end{table}

\subsection{\label{sec:C}Noise induced waves}
The results of LSA of a flat film (see Sec.~\ref{sec:lsa}) for the most unstable
wave number shown in Table~\ref{tab:wavelength}, confirm that a flat film is
unstable to long wave perturbations  for negative $D$'s. Although our base state is
not a flat film, there is clearly a possibility that numerical noise, which includes
long wave component, could grow in time and influence the results. As an example, we
consider again $D=-2.0$.  Similar results and conclusions can be reached for other
values of $D$.

\begin{figure}[th]
\centering
\includegraphics[scale=0.32]{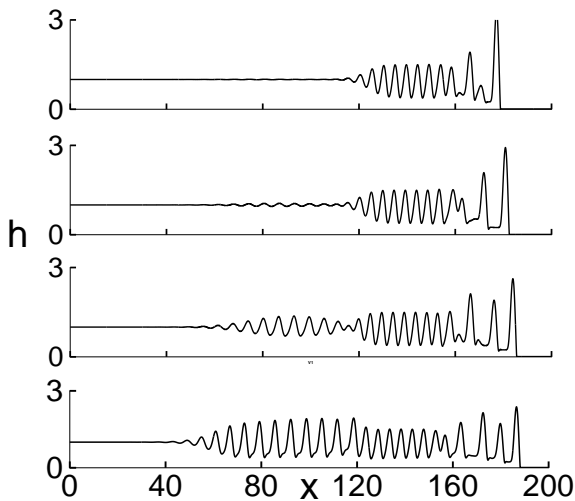}
\caption{$D=-2.0$. From top to bottom, $t=25$, $27$, $29$, $31$ (double precision).
The initial condition for this simulation is chosen to be a hyperbolic tangent with
contact line located at $x=150$. Also note the comparison between contact line induced
wave ($120 < x < 190$) and error-induced wave ($50 < x < 120$) (enhanced online).
}
\label{fig:LSA1}
\end{figure}

\begin{figure}[th]
\centering
\includegraphics[scale=0.32]{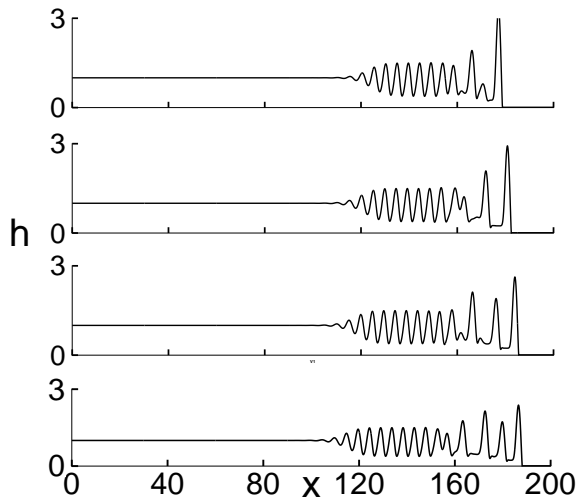}
\caption{$D=-2.0$. From top to bottom, $t=25$, $27$, $29$, $31$ (quadruple precision).
The initial condition is the same as in Fig.~\ref{fig:LSA1}.
}
\label{fig:LSA2}
\end{figure}

Let us first discuss expected influence of numerical noise. For $D=-2.0$, the LSA shows
that it takes $30$ time units for the noise of initial amplitude of $10^{-16}$ (typical
for double precision computer arithmetic) to grow to $10^{-2}$. LSA also shows that waves
with small amplitude should move with the speed $3$. That is, natural noise, which is
initially at $x=0$, should arrive to $x=90$ after $30$ time units. Figure~\ref{fig:LSA1}
illustrate this phenomenon. As $t$ approaches $30$, we see that the noise appears at about
$x=90$. Noise manifests itself through the formation of waves {\it behind} the contact line induced
waves which were already present for earlier times; see also the animation attached to
Fig.~\ref{fig:LSA1}. To further confirm that this new type of waves is indeed due to
numerical noise, we have also performed simulations using quadruple precision computer
arithmetic.  Figure~\ref{fig:LSA2} shows the outcome: with higher precision, the noise
induced waves are absent, as expected.  We note that in order to be able to clearly
identify various regimes, we take $x_f = 150$  in Figs.~\ref{fig:LSA1} and~\ref{fig:LSA2},
so that no influence of the boundary condition at $x=0$ is expected.

In Fig.~\ref{fig:LSA1} ($t=31$), we can clearly distinguish  between the waves induced by
contact line $(120 < x < 190)$, and the `natural waves' induced by noise $(50 < x < 120)$.
The main difference is the wavelength. The contact line induced waves have its specific
wavelength, $2\pi/k_l$, while the noise induced ones are characterized by a wavelength,
$\lambda$, corresponding very closely to the mode of maximum growth, $\lambda \approx {2 \pi/k_m}$,
obtained using LSA.  This can be clearly seen by comparing the numerical results shown in
Fig.~\ref{fig:LSA1} ($t=31$) with the LSA results given in Table~\ref{tab:wavelength}.

To summarize, the evolution of the wave structure in the {\it Type 2} and {\it Type 3}
regimes proceeds as follows.  First one sees formation of contact line induced waves,
characterized by relatively short wavelengths (compared to what would be expected
based on the LSA of a flat film).  Depending on the value of $D$, one may also see
formation of solitary-looking waves immediately following the capillary ridge. At
some later time, these waves are followed by noise-induced one. These three types
of waves are all presented in Fig.~\ref{fig:LSA1}. Then, at even later times, when the
waves cover the whole domain and interact with the $x=0$ boundary, the final wave
pattern forms, as illustrated for $D=-2.0$ by Fig.~\ref{fig:D_compare}. In Conclusions
we discuss briefly under which conditions these waves may be expected to be seen in
physical experiments.

{\bf Remark I.}
Finally, one may wonder why the traveling wave solution for $D=-1.0$ shown in
Fig.~\ref{fig:d10} remains stable for such a long time.  Recall that the LSA predicts
that natural noise with amplitude $10^{-16}$ should grow to $10^{-2}$ in $130$ time units,
while our numerical result shows that flat film is preserved even for $t=160$. The
reason is the domain size. It takes approximately $66$ time units for noise to travel
across the domain (moving with the speed equal to 3),  and the noise can only grow
from $10^{-16}$ to $10^{-9}$ during this time period for $D=-1$. This is why we do not
see the effect of noise for small $D$'s.

{\bf Remark II}
We have used LSA of a flat film (therefore, ignoring contact line presence) to
predict the evolution of the {\it size} of the region covered by waves in Sec.~\ref{sec:B}.
However, in order to understand the {\it properties} of the waves that form in this region,
one has to account for the presence of a front, as discussed in Sec.~\ref{sec:C}. In our
simulations, we are able to tune the influence of noise on the results (vis.
Fig.~\ref{fig:LSA1} versus Fig.~\ref{fig:LSA2}). In physical experiments, these
two effects will quite possibly appear together.

\subsection{Relation of physical quantities with non-dimensional parameter $D$}

It is useful to discuss the relation between the nondimensional parameter $D$
in our model, Eq.~(\ref{eq:dif-g}), and physical quantities.  In particular, we recall
that there are two quantities, $h_0$ (film thickness) and $\alpha$ (inclination angle)
which can be adjusted in an experiment, and here we discuss how variation of
each of these modifies our governing parameter and the results.  We also relate
$D$ to the fluid flux and the Reynolds number.

The velocity scaling in Eq.~(\ref{eq:dif-g}) can be expressed as
\[ U=\frac{\rho g}{3\mu}h^2_0\sin\alpha. \]
Therefore the parameter $D$ can be written as
\begin{equation} \label{D1} D = \left(\frac{\rho g}{\gamma}\right)^{1/3} h^{2/3}_0\frac{\cos\alpha}{(\sin\alpha)^{2/3}}. \end{equation}
In our simulations, the flux $Q$ in the $x$-direction is been kept constant and
equals to $1$. The dimensional flux is
\begin{equation}
\label{Q1} Q=1\cdot h_0\cdot U = \frac{\rho g}{3\mu}h^3_0\sin\alpha.
\end{equation}
Reynolds number can be expressed as
\begin{eqnarray}
\label{Re}Re = \frac{\rho U \ell}{\mu} = \frac{(\rho^{5} g^2 \gamma^2)^{1/3}}{3 \mu^2}
h_0^{7/3} (\sin\alpha)^{2/3}.
\end{eqnarray}
We note that there is no contradiction in considering $Re$, although inertial
effects were neglected in deriving the formulation that we use. The present
formulation is valid for $Re = O({1/\epsilon})$ or smaller, where $\epsilon \ll 1$ is the
ratio of the length scales in the out-of-plane and in-plane directions~\cite{Acheson}.
Considering the influence of $Re$ for this range is permissible.

\begin{table}[th]
\centering
\tabcolsep=8pt
\begin{tabular}[t]{ccccccc}
 & \multicolumn{2}{c}{$\alpha$ fixed} & & & \multicolumn{2}{c}{$h_0$ fixed}\\
\hline
 & $D>0$ & $D<0$ & & & $D>0$ & $D<0$\\
\hline
$|D|$ & $\uparrow$ & $\uparrow$ & & $|D|$ & $\uparrow$ & $\uparrow$\\
$h_0$ & $\uparrow$ & $\uparrow$ & & $\alpha$ & $\downarrow$ & $\uparrow$\\
$Q$ & $\uparrow$ & $\uparrow$ & & $Q$ & $\downarrow$ & $\downarrow$\\
$Re$ & $\uparrow$ & $\uparrow$ & & $Re$ & $\downarrow$ & $\downarrow$\\
$U$ & $\uparrow$ & $\uparrow$ & & $U$ & $\downarrow$ & $\downarrow$\\
\hline
\end{tabular}
\caption{Relation of the parameter $D$ to other parameters for fixed contact angle,
$\alpha$ (left), and for fixed film thickness, $h_0$ (right).
The up arrow $\uparrow$ means an increase and down arrow
$\downarrow$ means a decrease.
}
\label{tab:D}
\end{table}

The relation between $D$ and relevant physical quantities is shown in Table~\ref{tab:D}.
For fixed inclination angle an increase of the magnitude of $D$ is equivalent to an
increase of the film thickness, flux and Reynolds number. On the other hand, for fixed
film thickness, i.e., $h_0=$constant, raising the magnitude of $D$ leads to lower flux
and Reynolds number, and the inclination angle approaches horizontal. We can use this
connection to relate to the experimental results of Alekseenko et al. (see Fig. 11
in~\cite{alekseenko96}). They performed the experiments  with fixed inclination angle
and increasing the flux, which corresponds to an increase of the magnitude of $D$ in our
case. Our Fig.~\ref{fig:wave} shows that the trend of our results is the same as in the
above experiments. In addition, the results in~\cite{indeikina97} suggest that further
increasing of the flux leads to pinch-off, consistently with our results, since for
$D < -3.0$, numerics suggest that lubrication assumption is not valid.

Finally, one should recall that lubrication approximation is derived under the
condition of small slopes, which translates to
\begin{equation}
\left( \frac{h_0\sqrt{\sin\alpha}}{a} \right)^{2/3} \ll 1,
\label{D2}
\end{equation}
if nondimensional slopes are O$(1)$, see e.g.~\cite{siam}; here
$a= \sqrt {\gamma/\rho g}$ is the capillary length. In addition, by combining the
above lubrication limit with Eq.~(\ref{D1}), one gets the following condition
(see also~\cite{hom91}):
\begin{equation}
\label{D3}
|D| < |\cot\alpha|.
\end{equation}
Therefore, for a given $D$, there exists a range of inclination angle for which the
thin film model, Eq.~(\ref{eq:dif-g}), is valid. Table~\ref{tab:rang} shows this range for some
values of $D$.

\begin{table}[th]
\centering
\renewcommand{\tabcolsep}{1cm}
\begin{tabular}[t]{ccc}
\hline
$D$ &  $\alpha_c$ \\
\hline
-1.0 & $135^o$ \\
-1.5 & $147^o$ \\
-2.0 & $154^o$ \\
-3.0 & $162^o$ \\
\hline
\end{tabular}
\caption{
$\alpha_c$ is the inclination angle at which lubrication
theory ceases to be formally valid.}
\label{tab:rang}
\end{table}

\section {Conclusions}
In this paper we report numerical simulation of thin film equation~(\ref{eq:dif-g})
on inverted substrate. It is found that by changing a single parameter $D$, one can
find three different regimes of instability. Each regime is characterized by different
type of waves. Some of these waves show similar properties as the ones observed
in thin liquid films with periodic forcing. However, in contrast to those waves
produced by perturbations at the inlet region, our instability comes from the front.
We find that the presence of a contact line leads to free surface instability
without any additional perturbation. According to linear stability analysis,
we know that for negative $D$, the model problem,
Eq.~(\ref{eq:dif-g}), is unstable in the sense that any numerical disturbance
grows exponentially in time. However, we can also take advantage of the stability
analysis to separate the instability caused by noise and any other sources.

Finally, we may ask about experimental conditions for which the waves discussed
here can be observed. As an example, consider polydimethylsiloxane (PDMS), also
known as silicon oil  (surface tension: $21 dyn/cm$; density: $0.96 g/cm^3$), and
discuss the experimental parameters for which the condition $|D|<3.0$ is satisfied.
For $\alpha=170^o$ (the value used in~\cite{alekseenko96}), the thickness should be
less than $1.4 mm$. Table~\ref{tab:h} gives the values for this, as well as for some
other $D$'s. However, one should recall that Eq.~(\ref{D3}) shows that our model is
formally valid only up to a certain $D$ for a given inclination angle $\alpha$.
In addition, one should be aware that the use of lubrication approximation is easier
to justify for inclination angles further away from the vertical.

\begin{table}[th]
\centering
\renewcommand{\tabcolsep}{.4cm}
\begin{tabular}[t]{ccc}
\hline
 $\alpha$ &  $D$ & $h_0$($mm$)\\
\hline
$150^o$ & -1.0 & 0.93\\
 & -1.5 & 1.7\\
\hline
$170^o$ & -1.0 & 0.27\\
 & -2.0 & 0.75\\
 & -3.0 & 1.40\\
\hline
\end{tabular}
\caption{$h_0$ is the film thickness defined by Eq.~(\ref{D1}).}
\label{tab:h}
\end{table}

{\it Acknowledgments.}  The authors thank Linda J. Cummings and Burt Tilley for useful
comments.  They also acknowledge very useful input from anonymous referee leading to the
material presented in Sec.~\ref{sec:B}. This work was partially supported by NSF grant
No. DMS-0908158.

\appendix
\section{Evolution of small perturbations}
Equation~(\ref{eq:1D}) is a strongly nonlinear PDE and, to our knowledge, has no
analytical solutions. In this appendix, we present two analytical approaches which
consider evolution of small perturbations from a base state within linear approximation.
While these results are useful for the purpose of verifying numerical results, they also
provide a very useful insight into formation and evolution of various instabilities
discussed in this work.

\subsection{\label{sec:tws}Traveling wave solution}
Setting $s=x-Ut$ in Eq.~(\ref{eq:1D}), a traveling wave $H(s)=h(x,t)$ must satisfy
\begin{equation}
-UH+[H^3(H'''-DH'+1)]=c.\label{eq:traveling_wave}
\end{equation}
Imposing the conditions $H\rightarrow 1$ as $s\rightarrow -\infty$, and $H\rightarrow b$
as $s\rightarrow \infty$, we find $U=1+b+b^2$, $c=-b-b^2$~\cite{Troian,BB97}.  The
traveling wave speed, $U$, is useful for verifying whether or not the numerical
result is a traveling wave solution.

\begin{figure}[h]
\centering
\includegraphics[scale=0.3]{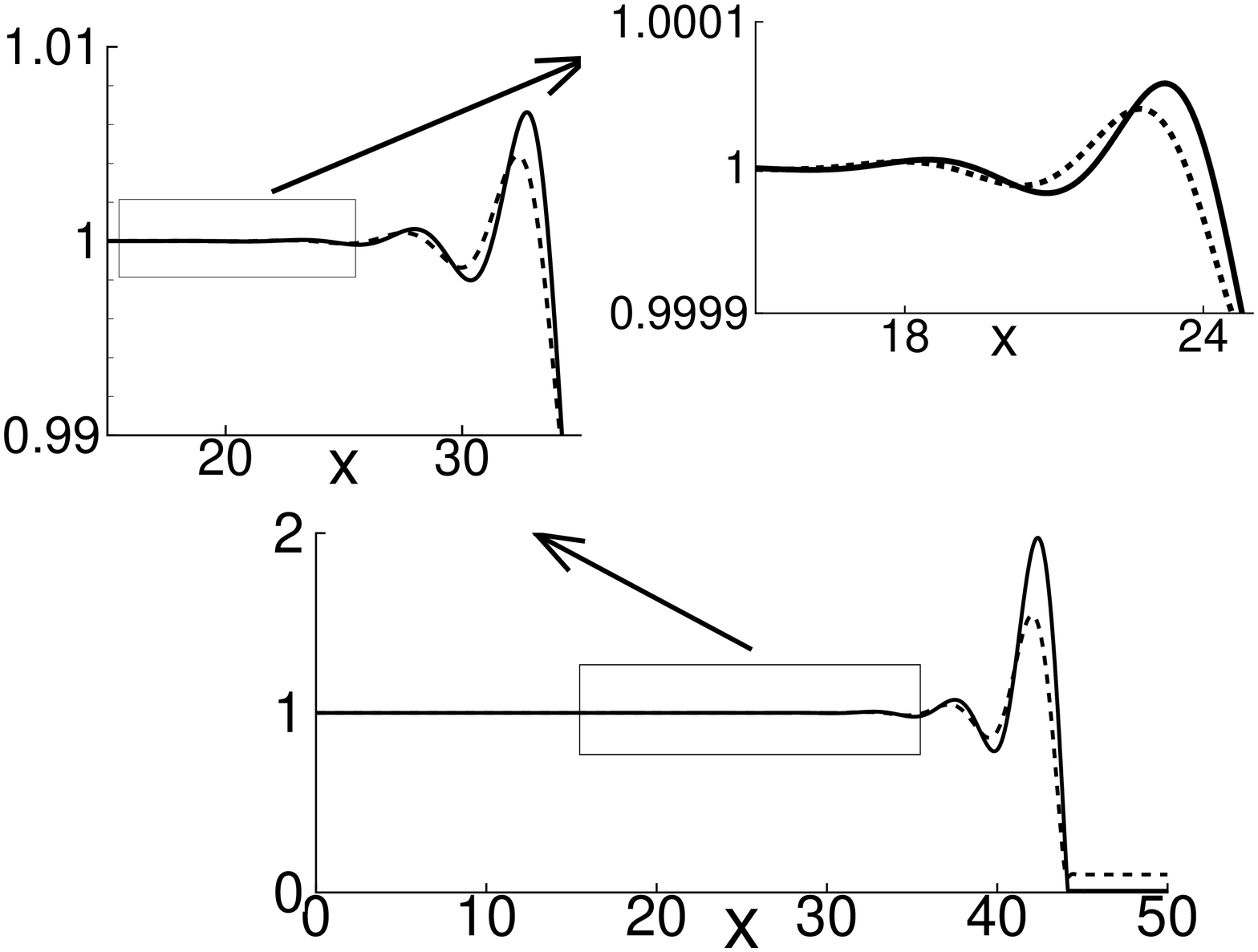}
\caption{Traveling wave solution of $D=-1$ case at three different scales with
precursor thickness $b=0.01$ (solid) and $b=0.1$ (dashed).  Note that the precursor
thickness only changes amplitude of the traveling wave profile but not the
wavelength.  The arrows point to the zoomed-in regions.}
\label{fig:tail_tw}
\end{figure}

Figure~\ref{fig:tail_tw} shows a typical profile of the traveling wave solution for $D=-1$.
A capillary ridge forms behind the fluid front, similarly as for the flow down a vertical or
inclined ($D>0$) substrate.
We also find that there exists a long oscillatory region behind the capillary ridge.
To analyze this 'tail', we expand Eq.~(\ref{eq:traveling_wave}) around the
base state, $H\equiv 1$, and consider the evolution of a small perturbation of the form
$exp(q s)$, where $q=q_r+iq_i$. We find
\begin{eqnarray}
-8q_r^3+2Dq_r+2-b-b^2 &=& 0,\label{tail_1}\\
q_i^2+D &=& 3q_r^2.\label{tail_2}
\end{eqnarray}
Table~\ref{tab:tail} shows the only positive root for $q_r$, for a set of D's. The
positivity of this root signifies that the amplitude of the tail decays exponentially
in the $-x$ direction, as also suggested by the insets of Figure~\ref{fig:tail_tw}.
Furthermore, as shown in Table~\ref{tab:tail}, $q_r$ decreases for more negative $D$'s,
meaning that the tail is longer for these $D$'s.   This table also shows the imaginary part
of $q$; we see an increase of its magnitude as $D$ becomes more negative, suggesting
shorter and shorter wavelengths in the tail.
Tail behavior is very useful for computational reasons.
For example, to solve Eq.~(\ref{eq:traveling_wave}) by shooting method, one can
evaluate suitable shooting parameters through Eqs.~(\ref{tail_1},\ref{tail_2}).

\begin{table}[h]
\centering
\renewcommand{\tabcolsep}{.6cm}
\begin{tabular}[t]{ccc}
\hline
 $D$ &  $q_r$ & $q_i$\\
\hline
   0 & 0.63 & 1.09\\
-1.0 & 0.50 & 1.32\\
-2.0 & 0.38 & 1.56\\
-3.0 & 0.30 & 1.81\\
\hline
\end{tabular}
\caption{Properties of the damped oscillatory region (tail) behind the capillary ridge.}
\label{tab:tail}
\end{table}

Figure~\ref{fig:tail_tw} also shows the effect of precursor thickness on traveling wave
solution. It is found that while the precursor thickness changes the height of the capillary
ridge, it has almost no effect on the wavelength of the tail.

\subsection{\label{sec:lsa}Linear stability analysis}
Another approach to analyze the stability of a flat film, is classical linear stability
analysis.  Assume $h=1+\xi$ where $\xi\ll 1$. Eq.~(\ref{eq:1D}) can be simplified to the
leading order as
\begin{equation}
\xi_t + \xi_{xxxx} -D \xi_{xx} + 3\xi_{x}=0.\label{linear}
\end{equation}
By putting $\xi\sim \exp i(kx-\omega t)$, where $\omega=\omega_r+i\omega_i$, we obtain
the dispersion relation
\begin{equation}
-i(\omega_r+i\omega_i) + k^4 + Dk^2 + 3 i k=0,\label{dispersion_linear}
\end{equation}
hence
\begin{equation}
\omega_r=3k, \quad \omega_i=-(k^4+Dk^2)=-(k^2+D/2)^2+D^2/4.
\end{equation}
As a result, for non-negative $D$'s, a flat film is stable under small perturbations. For
negative $D$'s, it is unstable for the perturbations characterized by sufficiently large
wavelengths. The critical wave number $k_c=\sqrt{-D}$, and the perturbation with wave
number $k_m=\sqrt{-D/2}$ has the largest growth rate. Besides, the speed of a linear wave
is $3$, and it is significantly larger than the traveling wave speed, $U$. As discussed
in the main body of the text, this speed is very important to identify the waves
induced by natural noise.

In addition, one should note that the maximum growth rate increases as inclination angle
$\alpha$ goes from $\pi/2$ to $\pi$.  Particularly, in the limiting case $\alpha\rightarrow\pi$
(hanging film), the growth rate is exactly the same as for thin film Rayleigh-Taylor
instability (e.g.,~\cite{Burgess01}).  Note that the scaling used in present work is not
appropriate for $\alpha \rightarrow \pi$; to establish the result for this case, one should
consider a different scaling, or the dimensional formulation of the problem.

\bibliographystyle{unsrt}
\bibliography{films}

\end{document}